\documentclass[preprint,groupedaddress,nofootinbib]{revtex4-1}
\usepackage{graphicx}    
\usepackage{dcolumn}    
\usepackage{bm}           
\usepackage{latexsym}
\usepackage{amssymb}

\begin{document}

\title{The Size of Compact Extra Dimensions from Blackbody Radiation Laws} 
\author{Ramaton Ramos}
\email{ramaton@gmail.com}
\affiliation{Centro Brasileiro de Pesquisas F\'{\i}sicas, Rua Dr. Xavier Sigaud 150, 
Urca, RJ 22290-180 -- Brazil}
\author{Henrique Boschi-Filho}
\email{boschi@if.ufrj.br}
\affiliation{Instituto de F\'{\i}sica, Universidade Federal do Rio de Janeiro, 
Caixa Postal 68528, RJ 21941-972 -- Brazil}

\date{\today}
 
\begin{abstract}
In this work we generalize the Stefan-Boltzmann and Wien's displacement laws for a 
$D$-dimensional manifold composed by 4 non-compact dimensions and $D-4$ 
compact dimensions, $ {\mathbb{R}}^{1,3} \times {\mathbb{T}}^{D-4} $. The 
electromagnetic field is assumed to pervade all compact and non-compact dimensions. 
In particular, the total radiated power becomes $ R(T) = \sigma_B T^4 + \sigma_D (a) 
\, T^D $, where $a$ is the size of the compact extra dimensions. For $D=10$, predicted 
from String Theory, and $D=11$, from M-Theory, the outcomes agree with available 
experimental data for $a$ as high as $ 2 \cdot 10^{-7}$m. 
\end{abstract}

\maketitle

\section{Introduction}

Interesting proposals to solve open problems in particle physics, like the Hierarchy 
issue \cite{hierextrad,hierarchymass}, and in cosmology, like the dark matter and 
dark energy, involve spacetimes with dimensions different from the usually accepted 
four. The original proposal of a model with one extra dimension was done by Kaluza 
and Klein in the 1920's to unify gravity, described by the general theory of relativity, 
with Maxwell's electromagnetism. In their model, the fifth dimension is compactified 
on a circle of a given radius, while the other four non-compact dimensions are 
identified with our usual space-time \cite{Dine:2007zp}.

String theory is presently the most viable candidate for unifying gravity 
with the other fundamental interactions described by quantum field 
theories and requires a 9+1 dimensional spacetime. String theory is also 
related to M-theory defined in 10+1 dimensional spacetime \cite{Becker:2007zj}.
Usually, the extra dimensions are supposed to be compact, with 
a compactification parameter that might be related to the Planck scale. 

About a decade ago, different models have been proposed to deal with the size 
of extra dimensions, keeping the extra dimensions compact 
\cite{ArkaniHamed:1998rs,Antoniadis:1998ig} or non-compact \cite{Randall:1999}. 
Many predictions of these and other models are being tested in the ongoing LHC 
experiments \cite{LHC,radion}, with hadronic beams colliding now at 7TeV and in the 
forthcoming years at 14TeV. 

Another way to test the predictions of string/M-theory is to look at the physics of 
low energy processes. It is expected that, in some way, string/M-theory shall reproduce 
the physics that one can observe in scales much lower than the Planck scale.

Blackbody radiation is very well described by Planck's law which implies, and explains, 
the Stefan-Boltzmann and Wien laws. In particular, the Stefan-Boltzmann law predicts 
that the electromagnetic radiated power of a blackbody is $R(T) = \sigma_{_B}\,T^4$, 
where $\sigma_{_B}$ is the Stefan-Boltzmann constant.  

Recently, it was pointed out that the blackbody radiaton laws should be corrected if 
the radiation is confined into a cavity \cite{Garcia-Garcia2008}. On the other side, it 
has been noticed that the blackbody radiation laws should depend on the (flat and 
non-compact) spacetime dimensions $D$ \cite{Cardoso:2005cd,Alnes:2005ed} such 
that the Stefan-Boltzmann law would be modified to $R(T) = \sigma_{_D} \, T^D$ 
with $\sigma_{_D}$ = constant, in constrast with observed data. 

The main result of this work is the blackbody radiated power $ R(T) = \sigma_B T^4 + 
\sigma_D (a) \, T^D $ for a spacetime with four non-compact dimensions and $D-4$ 
compact dimensions. This outcome reproduces, at low temperatures, the well-known 
Stefan-Boltzmann law and also the calculated relation for a $D$-dimensional spacetime 
\cite{Cardoso:2005cd,Alnes:2005ed} in the high temperature regime, 
as we are going to show in the following sections. 

\vspace{.2cm}
This article is organized as follows: In section 2 we present a brief account of the 
blackbody concept and introduce our approach. Section 3 is routed to the generalization 
of the Stefan-Boltzmann and Wien's displacement laws along with their peculiarities 
for a $ {\mathbb{R}}^{1,3} \times {\mathbb{T}}^{D-4} $ spacetime.  
In section 4 we consider recent high-temperature blackbody experiments in order to 
calculate bounds on the size of the considered compact dimensions for String theory 
($D$=10) and M-theory ($D$=11). The final section is devoted to general comments 
and closing remarks.

\section{Cavity Radiation}

A blackbody is defined as a body with a rich energy spectrum, capable of exciting all 
frequencies of light by thermalization. As consequence, all blackbodies emit thermal 
radiation with the same spectrum. For a current review on the matter, see \cite{Leq}.

In order to study its properties one conveniently takes a small bidimensional orifice 
connecting an isothermal enclosure to its outside as a blackbody surface (a more 
technical and conceptual discussion on approximating a blackbody for a blackbox can 
be found in \cite{Smerlak}). Here we consider that the blackbody is immersed in a 
$D$-dimensional spacetime, $ {\mathbb{R}}^{1,3} \times {\mathbb{T}}^{D-4} $. 

The electromagnetic radiation inside the enclosure is assumed to be composed of 
standing waves. Choosing a system of orthogonal coordinates with origin at one of 
the enclosure's vertices, we take for the sake of simplicity \cite{comment1} ${\ell}$ 
as the length of the edges parallel to the non-compact axes $x_i$ ($i=1,2,3$) and 
$a$ as the length of the edges related to the compact axes $x_j$ ($j=4,...,D-1$), 
so the $i$-th and $j$-th components of the electric field ($l=i,j$) are given by 
\begin{equation}
E_l(x_l,t) = E_{0,l} \sin({{\rm k}_l}x_l) \, e^{-{\rm i} \omega t} \ .  \label{El}
\end{equation} 

The $i$-th electric field components satisfy Dirichlet boundary conditions 
$E_i(0,t)=E_i(\ell,t)=0 \, $, for which there is no surface currents on the isothermal 
cavity walls, while the $j$-th components satisfy periodic boundary conditions 
$ E_j({x}_j,t) = E_j({x}_j+a,t) \, $, regarding the compactness of its corresponding 
dimensions. Thus
 \begin{equation}\label{kikj}
{\rm k}_i = \frac{\pi}{\ell} \, n_i \ , \hspace{1.8cm} 
{\rm k}_j = \frac{2\pi}{a} \, n_j \ ,
\end{equation}
where $n_i, n_j = 0,1,2,3,...$ represent the possible modes of vibration.

Considering the standing waves like components of a plain electromagnetic one with 
wave-vector ${\vec {\rm k}}$ and since the frequency $ \nu = |\vec {\rm k}|c / 2\pi $, 
we get for each mode $(n_i, n_j)$
\begin{equation}\label{ni}
  \nu_{ij} \equiv \nu(n_i,n_j) = \, \frac{c}{a} \sqrt{ \frac{a^2}{4{\ell}^2} 
       \sum_{i=1}^{3}{n_i}^{2} + \sum_{j=4}^{D-1}{n_j}^{2} } \;\;.
\end{equation} 

Making use of Bose-Einstein statistical prescription and accounting two helicity states 
related to the propagative aspect of the photons associated with the standing waves, 
since we do not consider polarization along the compact dimensions, the energy density 
inside the isothermal enclosure maintained at temperature $T$ is 
\begin{equation}\label{denserg}
{\rho}(T) = \frac{2}{V} \sum_{n_i,n_j} 
                    \frac{h\nu_{ij}}{e^{h\nu_{ij}/kT}-1} \ ,
\end{equation}
with $ V={\ell}^3 $, the 3-dimensional cavity volume.

The energy density is proportional to the radiancy $R(T)$, the energy rate per unit 
area of the orifice. Considering geometric factors for which the emanated power 
propagates only through the 3 non-compact dimensions, since the other ones are 
compact, one gets 
\begin{equation}\label{RTro}
R (T) = \frac{c}{4} \, \rho (T) \ .
\end{equation}

The compact dimensions have the same length of their corresponding edges $a$, 
which is originally unknown, while the orthodox edges are taken to have length 
$\ell$ at the cm scale or higher. Thus, although $n_i$ and $n_j$ cover the same 
numerical range, their respective contributions to $\nu_{ij}$ may not be on the 
same foot. With this in mind, (\ref{denserg}) will be worked out for two separate 
cases, $\sum {n_j}^2=0 \,$ and $\sum {n_j}^2 \neq 0$. Namely
\begin{equation}\label{radiancy2}
R (T) = \frac{c}{2V} \left[ \sum_{n_j=0} + \sum_{n_j \neq 0}
            \right] \frac{h \nu_{ij}}{e^{h\nu_{ij}/{kT}}-1} \ .
\end{equation}

\section{Generalizing Stefan-Boltzmann and Wien's displacement laws}

The first sum is a single one in $n_i$. Considering $\nu_{_0} \equiv \nu(n_i,0)$ 
and taking this sum by an integral, one gets
\begin{equation}
R_4(T) \equiv \frac{c}{2V} \sum_{n_j=0} \frac{h\nu_{_0}}
           {e^{h\nu_{_0}/{kT}}-1} 
           = \int_{0}^{\infty} R_4(T,\nu_{_0})\,d\nu_{_0}
\end{equation}
where the spectral radiancy $R_4(T,\nu_{_0})$ for this range of modes is 
\begin{equation}\label{R4nu}
R_4(T,\nu_{_0}) = \frac{2 \pi h}{c^2} 
                            \frac{{\nu_{_0}}^3}{e^{h \nu_{_0}/kT}-1} \,.
\end{equation}
 
Making the variable change $ z_{_0} = h\nu_{_0} / kT $ one integrates (\ref{R4nu}) 
arriving at
\begin{equation}
R_4(T) = \frac{2\pi c}{{(hc)}^3} {(kT)}^4 \int_{0}^{\infty} 
\frac{z_{_0}^3}{e^{z_{_0}}-1}dz_{_0} \ ,
\end{equation}
with the integral being expressed by the mathematical identity 
\begin{equation}\label{zetagama}
\int_{0}^{\infty}\frac{z^d}{e^z-1}\,dz = \Gamma (d+1)\,\zeta (d+1) \ .
\end{equation}

Taking $ d=3 $ in the above identity, we obtain for the radiancy contribution 
due to the radiation confined within the 3 non-compact dimensions 
\begin{equation}\label{R4}
R_4 (T) = \sigma_{_{B}} \, T^4 \ ,
\end{equation}
which is the well-known Stephan-Boltzmann law with
\begin{equation} 
\sigma_{_B} = \frac{2{\pi}^5 k c}{15} {\left( \frac{k}{hc} \right)}^3 = 
\, 5.67 \cdot 10^{-8} \, {\rm W} \, {\rm m}^{-2} {\rm K}^{-4} \, .
\end{equation}

\vspace{.1cm}
Now we consider the case $ \sum {n_j}^2 \neq 0 \, $ where both $n_i$ and $n_j$ 
contribute to the spectral radiancy. Taking the sum by an integral, with $\nu \equiv 
\nu(n_i,n_j)$, one gets
\begin{equation}\label{RDdef}
R_{_D}(T) \equiv \frac{c}{2V} \sum_{n_j \neq 0} 
           \frac{h\nu}{e^{h\nu/{kT}}-1} = 
           \int_{0}^{\infty} R_{_D}(T,\nu) \, d\nu 
\end{equation}
and the corresponding spectral radiancy is given by
\begin{equation}\label{RDnu}
R_{_D}(T,\nu) = \frac{{\Omega}_{(D-1)}}{2^{D-3}} 
                     \frac{a^{D-4}}{c^{D-2}} \, 
                     \frac{h\,{\nu}^{D-1}}{e^{h\nu/kT}-1} \ \ ,
\end{equation}
where ${\Omega}_{(d)} = 2 \, {\pi}^{d/2} / \, \Gamma (d/2)$ is the solid angle 
in a $d$-dimensional space. 

Making the variable change $ z = h\nu / kT $ one gets
\begin{equation}
R_{_D}(T) = \frac{{\Omega}_{(D-1)}}{2^{D-3}} 
               \frac{h\,a^{D-4}}{c^{D-2}} {\left( \frac{kT}{h} \right)}^D 
               \int_0^{\infty} \frac{z^{D-1}}{e^z-1} \, dz \ ,
\end{equation}
and using (\ref{zetagama}) the radiancy contribution $R_{_D}(T)$ due to these modes is 
\begin{equation}\label{RD}
R_{_D} (T) = \sigma_{_D}(a) \, T^D \ ,
\end{equation}
with 
\begin{equation}\label{sigmaD}
\sigma_{_D}(a) = 4 \, {\Omega}_{(D-1)} k c \, {\left( \frac{k}{2hc} \right)}^{D-1} 
         \Gamma (D)\,\zeta (D) \ a^{D-4} \ .
\end{equation}

\vspace{.1cm}
Grouping the computed radiancy contributions one gets for the total blackbody energy 
rate per unit area, 
\begin{equation}\label{RTotal}
R(T) = \sigma_{_B} \, T^4 + \sigma_{_D} (a) \, T^D \ ,
\end{equation}
which can be understood as the generalized Stefan-Boltzmann law for spacetimes with 
the $4$ usual non-compact dimensions and $D-4$ compact extra dimensions. 

The nature and validity of the preceding outcome is closely related with the 
temperature of the blackbody in question. As the reader can check directly via (\ref{ni}) 
and (\ref{denserg}), the $\nu$ contributions to $R(T)$ due to the compact extra 
dimensions are of the order of 
\begin{equation}
\frac{c}{2V} \frac{hc}{a} {(e^{hc/kaT}-1)}^{-1} \ \ \ \textrm{or lower.}
\end{equation} 
Thus, for low temperatures compared with the inverse of the compact dimensions size, 
$ T \ll {hc}/{k}{a} $, the sum of these statistical terms amount to an insignificant 
quantity and $R(T)$ reduces to ${\sigma_{_B} T^4}$, which is inferred as well through 
(\ref{RTotal}) since 
\begin{equation}\label{devT}
\frac{\sigma_{_D}(a) \, T^D}{\sigma_{_B} T^4} = \frac{15 \, {\Omega}_{(D-1)}}
        {4 \pi^5} \, \Gamma (D) \, \zeta (D) {\left( \frac{k}{2hc} \, a T \right)}^{D-4} . 
\end{equation}
So, for low temperatures, the obtained generalization for the Stefan-Boltzmann law 
(\ref{RTotal}) reduces to its well known form (\ref{R4}).

On the other side, for sufficiently high temperatures, $ T \gg {hc}/{k}{a} $, compact 
and non-compact dimensions match up on equal footing from the perspective of the 
actual irradiative process. Then, in view of (\ref{devT}), generalized Stefan-Boltzmann 
law takes on its higher-dimensional character $R(T)=\sigma_{_D}(a)\,T^D$, which 
agrees with the results in \cite{Cardoso:2005cd,Alnes:2005ed}. 

Also through (\ref{devT}) one observes that, the greater the number of extra compact 
dimensions, the narrower will be the transition temperature gap between the 
$\sigma_{_{B}} T^4$ and $\sigma_{_D}(a)\,T^D$ domains. 

\vspace{.2cm}
To obtain the respective Wien's law we write the total wavelength radiancy as 
\begin{eqnarray}\label{Rlambda}
R(T,\lambda) &=& R_4(T,\lambda)+R_{_D}(T,\lambda) \ ,
\end{eqnarray}
where 
\begin{eqnarray}
R_4(T,\lambda) &=& \frac{2 \pi h c^2}{\lambda^5} \, 
                     {(e^{\frac{hc}{k T \lambda}}-1)}^{-1} \ ,
\end{eqnarray}
and 
\begin{eqnarray}
R_{_D}(T,\lambda) &=& \frac{\Omega_{(D-1)} h c^2}{2 \, \lambda^{D+1}} \, 
                     {\left( \frac{a}{2} \right)}^{D-4} 
                     {(e^{\frac{hc}{k T \lambda}}-1)}^{-1} \ .
\end{eqnarray}
Since for a given $T$ there is a $\lambda_{_m}$ for which $R(T,\lambda_{_m})$ 
is maximum, through $dR(T,\lambda)/d\lambda = \, 0$ one obtains 
\begin{equation}\label{gwien}
1 - e^{-{hc/kT\lambda_{_m}}} = \frac{hc}{k T \lambda_{_m}} \, 
                              \frac{1+\epsilon_{_D}(a,\lambda_{_m})}
                              {5+(D+1)\,\epsilon_{_D}(a,\lambda_{_m})} \ \, ,
\end{equation}
where $\epsilon_{_D} (a,\lambda) = {R_{_D}(T,\lambda)}/{R_4(T,\lambda)}$ 
is defined as the {\it wavelength radiancy relative deviation}, 
\begin{equation}\label{wavelengthdeviation}
\epsilon_{_D} (a,\lambda)  = \frac{\Omega_{(D-1)}}{4\pi}
                                      {\left(\frac{a}{2\lambda}\right)}^{D-4} \ .
\end{equation}

Eq. (\ref{gwien}) can be regarded as the generalized Wien's law for spacetimes with 
$D-4$ compact extra dimensions, once it relates the blackbody temperature $T$ with 
the wavelength $\lambda_m$ corresponding to the maximum value of the total 
wavelength radiancy $ R(T,\lambda)$.

Note that for wavelenghts much larger than the size of compact dimensions, 
$ \lambda_m \gg a $, (\ref{gwien}) reduces to $ 1 - e^{-x_0}= \frac 15 x_0 $ 
where $x_0\equiv  {h c} / {k \lambda_m T}$. Since $x_0=W(-5e^{-5})+5 \approx 
4.965$, where $W(z)$ is the Lambert function, this implies the usual Wien's 
displacement law, namely $ \lambda_m T = 2.897 \cdot 10^{-3} \rm m \, K $. 

On the opposite limit, for  wavelengths much smaller than the size of compact extra 
dimensions $\lambda_m \ll a $, (\ref{gwien}) reduces to 
\begin{equation}\label{lowgwien}
 1 - e^{-x_{_{D}}} = 
\frac{x_{_{D}}}{D+1} \quad \;; \quad x_{_{D}} \equiv {\frac{h c}{k \lambda_m T}}
\end{equation}
where $x_{_{D}}=W(-(D+1)e^{-(D+1)})+D+1$. This yields $\lambda_m T = 
x_{_{D}} hc/k$ and the generalized Wien's displacement law assumes its 
D-dimensional behavior.

\section{Bounds on D=10 and D=11 scenarios}

\begin{figure}[!ht]
\begin{center}
\includegraphics{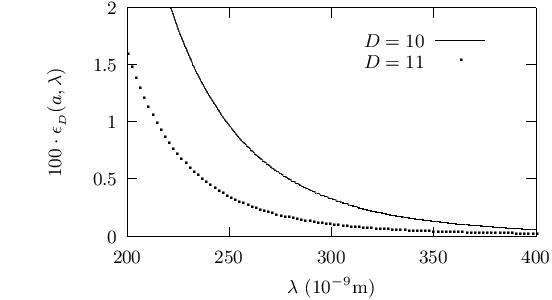}
\end{center}\caption{Wavelength radiancy relative deviation for $D=10$ and 
$D=11$ with respect to $D=4$, for $a = 2 \cdot 10^{-7}$m.}
\label{dev}
\end{figure}

Since every blackbody radiation measurement is performed within certain wavelength 
ranges, the upper bound for the size of compact dimensions can be calculated by 
fitting $\epsilon_{_D}(a,\lambda)\,$, for a given $D$, into the experimental uncertainty. 
Precise measurements 
\cite{codata2010,Sapritsky1995,Yoon2000,Friedrich2000,Yoon2003} performed at 
$\lambda \approx 250\,$nm and $T \approx 3000\,$K restrain the size of compact 
extra dimensions within our model to be not higher than $2 \cdot 10^{-7}$m for 
$D=10$ and $D=11$. In fig.(\ref{dev}), we plot the wavelength radiancy relative 
deviation $\epsilon_{_D} (a,\lambda)$ for these two situations. 

The generalized Stefan-Boltzmann law (\ref{RTotal})  for $D=10$ and $D=11$ are 
given respectively by (in S.I. units)
\begin{eqnarray}\label{T10}
R (T) &=& 5.67 \cdot 10^{-8} \, T^4 \, + \, 1.32 \cdot 10^7 a^6 \, T^{10} \, ,      
\\ \label{T11}
R (T) &=& 5.67 \cdot 10^{-8} \, T^4 \, + \, 3.94 \cdot 10^9 a^7 \, T^{11} \, .  
\end{eqnarray}

\begin{figure}[!ht]
\begin{center}
\includegraphics{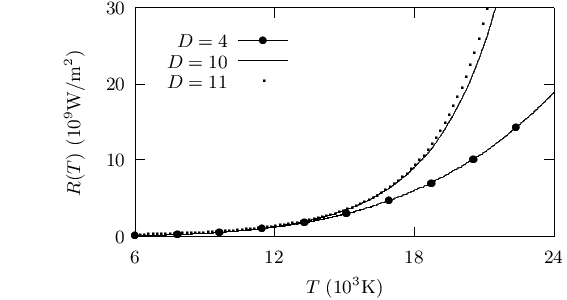}
\end{center}\caption{Generalized Stefan-Boltzmann law for $D=10$ and $D=11$  
compared with $D=4 \,$, for $a = 2 \cdot 10^{-7}$m.}
\label{sb}
\end{figure}

These expressions are plotted in fig.(\ref{sb}) and compared with the standard 
results. For the considered size of the compact extra dimensions the discrepancy 
between theses functions is just noticeable for $T > 10^4$K.

\vspace{.15cm}
Regarding the generalization of Wien's displacement law eq.(\ref{gwien}) for $D=10$ 
and $D=11$ one finds respectively ($\lambda = \lambda_{_m}$), 
\begin{eqnarray}\label{wien10}
1 - \exp\left({\frac{-\alpha}{\lambda T}}\right)=\frac{\alpha}{\lambda T} 
\, \frac{1+\beta{\left(\frac{a}{2\lambda}\right)}^6} 
      {5+11\beta{\left(\frac{a}{2\lambda}\right)}^6} \ ;  
\\ \label{wien11}
1 - \exp\left({\frac{-\alpha}{\lambda T}}\right)=\frac{\alpha}{\lambda T} 
\, \frac{1+\gamma{\left(\frac{a}{2\lambda}\right)}^7} 
      {5+12\gamma{\left(\frac{a}{2\lambda}\right)}^7} \ ,
\end{eqnarray}
where
\begin{eqnarray}
\alpha = \frac{hc}{k}\ \ , \ \ \beta = 2.36238 \ \ , \ \ \gamma = 2.02935 \ . \ \ \ 
\end{eqnarray}

\begin{figure}[!ht]
\begin{center}
\includegraphics{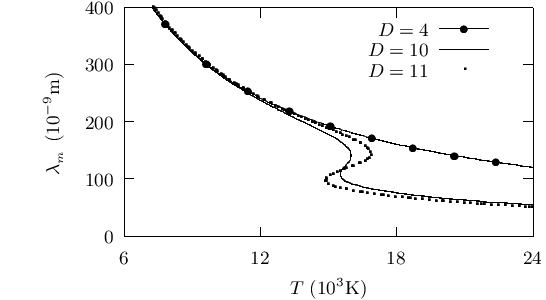}
\end{center}\caption{Generalized Wien's displacement law for $D=10$ and $D=11$, 
compared with $D=4 \,$, for $a = 2 \cdot 10^{-7}$m.}
\label{wien}
\end{figure}

The behavior of these generalized Wein's laws are plotted in fig.(\ref{wien}) and 
compared with the standard results. This graph also shows a clear signature of the 
extra dimensions for $T>10^4$K.

\vspace{.1cm}
An alernative setup that might relate String theory with M-theory regarding blackbody 
radiation laws can be achieved considering a space with the usual three non-compact 
dimensions and seven compact dimensions, six of which with length $a$ and the other 
one with length $b \ll a$. Following the same previous steps one can take
\begin{equation}
{\rm k}_i = \frac{\pi}{\ell}\,n_i \ , \hspace{1cm} 
{\rm k}_j = \frac{2\pi}{a}\,n_j \ ,  \hspace{1cm} 
\kappa = \frac{2\pi}{b} \, \eta \ ,
\end{equation}
so the total radiated power can be written as 
\begin{equation}\label{radiancySMtheory}
R (T) = \frac{c}{2V} \left[ \sum_{{\eta,n_j=0}} + 
            \sum_{\eta=0, n_j \neq 0} + \sum_{\eta \neq 0} \;\;
            \right] \frac{h\nu} {e^{\frac{h\nu}{k T}}-1} \ ,
\end{equation}
which leads to the respective blackbody radiation laws 
\begin{eqnarray}\label{RSM}
R(T) = \sigma_{_B}\,T^4 + \, \sigma_{_{10}} (a)\,T^{10} + 
\, \frac{b}{a} \, \sigma_{_{11}}(a) \, T^{11} \ , \\
1 - \exp\left({\frac{-\alpha}{\lambda T}}\right) = \frac{\alpha}{\lambda T} 
\frac{1+\beta{\left(\frac{a}{2\lambda}\right)}^6+
         \gamma \frac{b}{a} {\left(\frac{a}{2\lambda}\right)}^7} 
      {5+11\beta{\left(\frac{a}{2\lambda}\right)}^6+
         12\gamma \frac{b}{a} {\left(\frac{a}{2\lambda}\right)}^7} \ . 
\end{eqnarray}

Thus, the wavelength radiancy relative deviation in this case is given by
\begin{equation}
\epsilon_{_{SM}} (a,\lambda) = \beta{\left(\frac{a}{2\lambda}\right)}^{6}  
                          + \gamma \frac{b}{a}{\left(\frac{a}{2\lambda}\right)}^{7} \ .
\end{equation}
Assuming that there is a wavelength $\lambda = \lambda_{_L}$ for which $R_{10}
(T,\lambda)$ is preponderant over $R_4(T,\lambda)$ and $\frac{b}{a} R_{11} 
(T,\lambda)$ to ensure the desirable energy scale behavior, namely 
\begin{equation}
\beta{\left(\frac{a}{2\lambda}\right)}^{6} \approx \, 100 \ , \ \ 
\gamma \frac{b}{a}{\left(\frac{a}{2\lambda}\right)}^{7}  \leq  \,
\beta{\left(\frac{a}{2\lambda}\right)}^{6}/100 \ , 
\end{equation}
\begin{equation}
{\rm so} \hspace{.6cm} b \, \leq \, \frac{a}{100} \frac{\beta}{\gamma}
                      {\left(\frac{\beta}{100}\right)}^{1/6} \, ,  
\end{equation}
which provides an upper bound for $b$ with respect to $a \,$. 
So, for $a = 200 \,{\rm nm} $ one gets $ b \leq 1.2 \,{\rm nm} $.

\section{Conclusion and discussions}

In this work we obtained high energy corrections on the Blackbody Radiation Laws for 
a higher-dimensional scenario as well as an upper bound on the size of the considered 
compact extra dimensions, as initially proposed. The temperature for which deviations 
in the blackbody radiation becomes relevant is inversely proportional to the size of the 
compact dimensions, as can be seen in (\ref{devT}), while the wavelength for which 
deviations in the blackbody spectrum becomes important is directly proportional to the 
size of compact dimensions, as can be noted in (\ref{wavelengthdeviation}).

Regarding the validity of the approximation of sums by integrals used to reach the 
generalized blackbody radiation laws, particularly in (\ref{RDdef}), we point out 
that if the approximation is poor (for instance when $ a \ll hc/kT $) the obtained 
higher-dimensional terms are naturally negligible as seen directly from (\ref{radiancy2}).

Our approach not just yields the suitable $D$-dimensional blackbody behavior (as 
intended, to match the dimensional requirements of certain appealing theories) in 
the high energy regime, but is also compatible with our empirically presumed 
$4$-dimensional spacetime, once it reproduces the observed Stefan-Boltzmann 
and Wien's displacement law in our ordinary energy scale. 

The additive aspect of the generalized Stefan-Boltzmann and Wien's displacement laws, 
(\ref{RTotal}) and (\ref{gwien}) respectively, traces back to the temperature induced 
split structure of (\ref{radiancy2}), the key conceptual point of our model. 
Thus the blackbody temperature influences how the $D$-dimensional spacetime 
takes part in the blackbody radiation. In the low energy scale, it is seen that our 
${\mathbb{R}}^{1,3} \times {\mathbb{T}}^{D-4}$ spacetime behaves effectively 
as ${\mathbb{R}}^{1,3}$, while in the high energy scale ${\mathbb{R}}^{1,3} 
\times {\mathbb{T}}^{D-4}$ behaves effectively as ${\mathbb{R}}^{1,D-1}$. 

We also emphasize the significance of $\epsilon_{_D} (a,\lambda)$ in our approach, 
for this is the physical quantity through which an upper bound on the size of these 
compact extra dimensions can be estimated at experiments performed within short 
wavelengths as in \cite{Sapritsky1995,Yoon2000,Friedrich2000,Yoon2003}.

\section{Acknowledgments}

The authors appreciate interesting discussions with A.M. Garcia-Garcia, Nelson R.F. 
Braga, J.A. Helayel Neto and Konstantin Zioutas. The authors also acknowledge well 
observed points by the anonymous referee. The authors are partially supported 
by Capes and CNPq - Brazilian agencies.


\begin{thebibliography}{99}

\bibitem{hierextrad}
  G.~Burdman and A.~G.~Dias,
  JHEP {\bf 01} (2007) 041.

\bibitem{hierarchymass}
  Zhi-Qiang Guo and Bo-Qiang Ma, 
  JHEP {\bf 09} (2009) 091.

\bibitem{Dine:2007zp}
  For a review see, {\sl e. g.}, M.~Dine,
  ``Supersymmetry and string theory: Beyond the standard model,''
  Cambridge, UK: Cambridge Univ. Pr. (2007) 515 p.

\bibitem{Becker:2007zj}
  K.~Becker, M.~Becker, J.~H.~Schwarz,
  ``String theory and M-theory: A modern introduction,''
  Cambridge, UK: Cambridge Univ. Pr. (2007) 739 p.

\bibitem{ArkaniHamed:1998rs}
  N.~Arkani-Hamed, S.~Dimopoulos and G.~R.~Dvali,
  Phys.\ Lett.\  B {\bf 429}, 263 (1998)
  [arXiv:hep-ph/9803315].

\bibitem{Antoniadis:1998ig}
  I.~Antoniadis, N.~Arkani-Hamed, S.~Dimopoulos and G.~R.~Dvali,
  Phys.\ Lett.\  B {\bf 436}, 257 (1998)
  [arXiv:hep-ph/9804398].
  
\bibitem{Randall:1999}
  L.~Randall and R.~Sundrum,
  Phys.\ Rev.\ Lett.\  {\bf 83}, 3370 (1999); {\sl ibid.}  4690 (1999).

\bibitem{LHC}
  M.~Cicoli, C.~P.~Burgess and F.~Quevedo,
  JHEP {\bf 10} (2011) 119.

\bibitem{radion}
  H.~de~Sandes and R.~Rosenfeld,
Phys. Rev. D {\bf 85}, 053003 (2012).

\bibitem{Garcia-Garcia2008}
A M Garcia-Garcia, 
Phys. Rev. A {\bf 78} 023806 (2008). 

\bibitem{Alnes:2005ed}
  H.~Alnes, F.~Ravndal and I.~K.~Wehus,
  J.\ Phys.\ A  {\bf 40}, 14309 (2007).

\bibitem{Cardoso:2005cd}
  T.~R.~Cardoso and A.~S.~de Castro,
  Rev.\ Bras.\ Ens.\ Fis.\  {\bf 27}, 559 (2005).

\bibitem{Leq}
  R.~Lehoucq, 
  Eur. J. Phys. {\bf 32} (2011) 1495–1514.

\bibitem{Smerlak}
  M.~Smerlak,
  Eur. J. Phys. {\bf 32} (2011) 1143–1153.

\bibitem{comment1} {One could actually have taken ${\ell}_i \approx \ell$ and $a_j 
\approx a$ as the edges length, and the outcomes would be quite the same.}

\bibitem{codata2010} P. J. Mohr {\sl et. al.}, 
CODATA, 2010 
NIST http://physics.nist.gov/cuu/Constants/

\bibitem{Sapritsky1995} 
V I Sapritsky,
Metrologia, {\bf 32} (1995) 411-417.

\bibitem{Yoon2000} 
H W Yoon and C E Gibson,
Metrologia, {\bf 37} (2000) 429-432.

\bibitem{Friedrich2000} 
R Friedrich and J Fischer, 
Metrologia, {\bf 37} (2000) 539-542.

\bibitem{Yoon2003}
H W Yoon, C E Gibson and P Y Barnes,
Metrologia, {\bf 40} (2003) S172-S176.

\end{thebibliography}
\end{document}